\documentclass[12pt]{iopart}
\usepackage{graphicx}

\newcommand{\beq}{\begin{equation}}
\newcommand{\eeq}{\end{equation}}
\newcommand{\bea}{\begin{eqnarray}}
\newcommand{\eea}{\end{eqnarray}}
\newcommand{\abs}[1]{\vert#1\vert}
\renewcommand{\d}{{\rm d}}
\newcommand{\dpar}{\partial}
\renewcommand{\e}{{\rm e}}
\newcommand{\eps}{\varepsilon}
\newcommand{\eq}{{\rm eq}}
\newcommand{\g}{\gamma}
\newcommand{\imi}{{\rm i}}
\newcommand{\m}{{\rm m}}
\newcommand{\mean}[1]{\langle#1\rangle}
\newcommand{\prob}{{\rm Prob}}
\newcommand{\thr}{{\rm th}}
\newcommand{\w}{\widehat}
\newcommand{\B}{{\rm B}}
\newcommand{\D}{\Delta}
\newcommand{\F}{{\cal F}}
\newcommand{\R}{{\rm r}}
\eqnobysec

\begin{document}
\title[Dynamics at the angle of repose]
{Dynamics at the angle of repose:\\ jamming, bistability, and collapse}
\author{J M Luck\dag\ and Anita Mehta\ddag}

\address{\dag\ Service de Physique Th\'eorique\footnote{URA 2306 of CNRS},
CEA Saclay, 91191 Gif-sur-Yvette cedex, France}

\address{\ddag\ S N Bose National Centre for Basic Sciences, Block JD,
Sector 3, Salt Lake, Calcutta 700098, India}

\begin{abstract}
When a sandpile relaxes under vibration, it is known that its measured angle
of repose is bistable in a range of values bounded by a material-dependent
maximal angle of stability; thus, at the same angle of repose,
a sandpile can be stationary or avalanching, depending on its history.
In the nearly jammed slow dynamical regime, sandpile collapse
to a zero angle of repose can also occur, as a rare event.
We claim here that fluctuations of {\it dilatancy} (or local density)
are the key ingredient that can explain such varied phenomena.
In this work, we
model the dynamics of the angle of repose and of the density fluctuations,
in the presence of external noise, by means of coupled stochastic equations.
Among other things, we are able to describe sandpile
collapse in terms of an activated process, where an effective temperature
(related to the density as well as to the external vibration intensity)
competes against the configurational barriers created by the density
fluctuations.
\end{abstract}

\eads{\mailto{luck@spht.saclay.cea.fr},~\mailto{anita@bose.res.in}}
\maketitle

\section{Introduction}

The angle of repose~\cite{br} in a sandpile has been
a recurrent enigma for physicists, both in paradigmatic and in realistic terms.
Ideas of self-organised criticality~\cite{bak} used the angle of repose
as a paradigm -- it was rather appealing
to think that sandpiles relaxed to a marginally
stable state with a {\it unique} critical angle,
which resulted from a universal and extensive dynamics.
Interest in the angle of repose only increased
when it became obvious to most physicists that
this picture was too simplistic~\cite{sidrevmodphys}; in fact the angle
of repose is multiply-valued within a certain range,
and relaxational behaviour about it
reveals a great deal about intrinsic length
scales within the sandpile~\cite{mnd}.

We first summarise the relevant phenomenology.
Sandpiles formed by the deposition of grains
on surfaces have sides that are typically inclined
at a finite angle to the horizontal\footnote{Sandpiles
formed by other processes, e.g.~by drainage through a narrow
pore within a flat box, also have angles of repose
which depend on their history;
their description can also be included, with minor
modifications, in our formulation.}.
This is the angle of repose~$\theta_\R$: in practice what is seen
is that it can take a range of values before spontaneous
flow occurs, i.e., the sandpile becomes unstable to further deposition.
The limiting
value of the angle before avalanching occurs is known as the {\it maximal
angle of stability} $\theta_\m$.
The difference between these two angles,
which is characteristic of a given material~\cite{br},
is often referred to as the Bagnold angle~\cite{bagnold}.
Also, sandpiles show strong hysteresis\footnote{This is a result
of the `athermal' nature of sandpiles, i.e., a consequence of the fact that
grain sizes are too large for the ambient temperature
to have any effect on their dynamics.
Thus configurations that would be dissolved away by Brownian motion
in liquids or gases, survive in sandpiles.}: thus, depending
on its conditions of formation, a sandpile can either be stable or in motion
at any angle~$\theta$ such that $\theta_\R<\theta<\theta_\m$.
This {\it bistable} behaviour has been studied theoretically
and experimentally~\cite{bistability,daerr},
but its dynamical origin has not been
clearly explained so far, to the best of our knowledge.

Another enigma concerns the dynamics around the angle of repose.
Early experiments in the physics literature~\cite{sidprl} indicated that,
when subjected to a large vibration intensity,
the angle of repose of a sandpile would decay to zero logarithmically
with time: gently vibrated sandpiles, on the other hand, after
an initial logarithmic decay, remained `jammed'
at a finite angle of repose for experimentally observable times.
This led, in a theory~\cite{mnd}
that followed, to the suggestion that different mechanisms
were responsible for decay in the two dynamical regimes.
In the strongly vibrated sandpile, grains relaxed {\it independently} of each
other, and the angle of repose decayed logarithmically
to zero; for gently vibrated sandpiles, there was insufficient
inertial energy given to individual grains, and {\it collective}
dynamics were responsible for the `jamming' observed at finite angles.
The mean-field approach
adopted in that work was adequate to identify the different
rates of decay; a treatment of the fluctuations, needed
for example to explain bistability or jamming at the angle of repose,
was, however, beyond its scope.

In this work, our objectives are, firstly, to propose
a possible mechanism for observed bistable behaviour;
secondly, and equally importantly, to construct
a coherent picture, involving both fast and slow dynamical
modes, of relaxation at the angle of repose.
The present work is based on coupled stochastic equations
which are similar, but not identical, to those proposed earlier in~\cite{mnd}.
The exact analytical treatment to these equations will provide,
in the slow dynamical regime of interest, a description of the dual problems
of stability and collapse of the angle of repose.

We now summarise the main results of the present work.
Our basic picture is that fluctuations of local density
(especially when they occur coherently, in the limit of a small perturbation)
are the collective excitations responsible for stabilising the angle of repose,
and for giving it its characteristic width,
known as the {\it Bagnold angle}~\cite{bagnold}.
Density fluctuations can occur because of shape
effects~\cite{uscolumn} or friction~\cite{br,edwards98};
they are the manifestation in our model of
{\it dilatancy}, first observed by Reynolds~\cite{reynolds} in~1885.

Our theory below essentially
relates the Bagnold angle to the Reynolds' dilatancy.
In the absence of dilatancy, we suggest that
the angle of repose in a vibrated sandpile
would decay swiftly to zero\footnote{In liquids, for instance,
the absence of dilatancy is the reason
why liquid surfaces do not spontaneously sustain themselves
at a non-zero angle to the horizontal.}.
When dilatancy is present, density fluctuations
{\it add} to the value of the `bare' angle of repose.
In our theory, `out-of-equilibrium' density fluctuations
(such as those that might be found at the start of a shaking process)
generate the maximal angle of stability $\theta_m$; on the other hand,
asymptotic values of density fluctuations `equilibrated'
by vibrations at long times, give rise to the
the `typical' angle of repose $\theta_\R$.
The Bagnold angle, which is defined~\cite{bagnold} as the difference
\beq
\delta\theta_\B=\theta_\m-\theta_\R,
\label{ba}
\eeq
is thus found to be {\it the difference
between nonequilibrium and equilibrium values of the dilatancy
for a given material}.

Consider a sandpile that has relaxed to $\theta_\R$
in the presence of low noise; it
is now stabilised by an equilibrium value of the density fluctuations.
We ask the question: what is the probability, under these
circumstances, that the sandpile collapses
to a zero angle of repose?
The configurational landscape we are dealing with is that
of grains rather close to jamming; density fluctuations involve
small (typically intra-cage) displacements of grains about their
`equilibrium' positions in this very disordered network.
Typical configurational barrier heights for collapse under such
circumstances would be rather high, as they would involve
a {\it global} rearrangement of grains.
This is rather reminiscent of the situation close to the `dynamical transition'
in an earlier random graph model~\cite{johannes} of granular compaction;
in both cases, long-range correlations (corresponding to system-wide
density fluctuations) need to develop for an appropriate collapse
to occur\footnote{In the case of compaction, `collapse' corresponds
to the attainment of the dynamical transition via a collapse of excess
void space; here it corresponds to the collapse of the sandpile
to a zero angle of repose.}.
Clearly such events would be rare; while
they would clearly be facilitated by an increase in the external noise,
the effect of dilatancy merits more discussion.
An increase of dilatancy means that density fluctuations
are greater; the effect of disorder is greater and configurational landscapes,
rougher.
Since sandpile collapse requires global rearrangement of grains,
it will be less probable in rough (strongly dilatant) landscapes.
Our analytical results confirm these qualitative
observations; we find spontaneous collapse to be an {\it activated}
process, with an effective temperature~$\eps$ that depends on the ratio
of the external noise to the ambient value of the density fluctuations.

\section{The model}

The dynamics of the angle of repose $\theta(t)$ and of the density
fluctuations $\phi(t)$ are described by the following
coupled stochastic equations:
\bea
&&\dot\theta=-a\theta+b\phi^2+\D_1\,\eta_1(t),
\label{dy1}\\
&&\dot\phi=-c\phi+\D_2\,\eta_2(t).
\label{dy2}
\eea
The parameters $a$,~...,~$\D_2$ are phenomenological constants,
while $\eta_1(t)$, $\eta_2(t)$ are two independent white noises such that
\beq
\mean{\eta_i(t)\eta_j(t')}=2\,\delta_{ij}\,\delta(t-t').
\label{white}
\eeq

Similar equations, which involve spatial rather than temporal
coordinates, have been written to describe the orientational
structure of bridges in granular media; these have been extensively
analysed in concurrent work~\cite{bridges} and their results
are in good agreement with independent simulations.
Density fluctuations
are key to both phenomena; the angle of repose is stabilised by dilatancy,
while linear bridges grow at the expense of local density fluctuations.

From a technical viewpoint,
the linear equation~(\ref{dy2}) for $\phi(t)$ is
known as an Ornstein-Uhlenbeck equation~\cite{ou,vank},
whereas equation~(\ref{dy1}) for $\theta(t)$ is non-linear,
as it contains a quadratic coupling
to the Ornstein-Uhlenbeck variable $\phi(t)$.
In spite of this nonlinearity,
we have been able to find the
equilibrium state of the coupled equations~(\ref{dy1}),~(\ref{dy2})
analytically -- one of the very rare instances where this is possible.
In the Appendix, we show that the stationary Fokker-Planck equation
describing this equilibrium state
can be solved in closed form by means of a quadratic Ansatz~(\ref{ansatz}).
This ansatz only works for the very special kind of non-linearity here,
where~(\ref{dy1}) involves the {\it square}
of the Ornstein-Uhlenbeck variable $\phi$.
We have come across a similar quadratic Ansatz
for a problem with a quadratic coupling
between two stochastic variables~\cite{luczka}.

We now motivate our equations physically.
The first terms in~(\ref{dy1}) and~(\ref{dy2}) are suggested
on stability grounds:
neither the angle of repose nor the dilatancy are allowed
to be arbitrarily large for a stable system.
The second term in~(\ref{dy1}) affirms that dilatancy underlies the phenomenon
of the angle of repose; in the absence of noise,
density fluctuations constitute this angle\footnote{We write
a term proportional to $\phi^2$ on symmetry grounds --
we would expect it to depend on the magnitude rather than the sign
of density fluctuations.}.
The noise in~(\ref{dy1}) represents external vibration,
while that in~(\ref{dy2}) is a version
of the Edwards compactivity~\cite{sam}, related as it is
to purely density-driven effects.
In earlier work~\cite{mnd},
these were related via fluctuation-dissipation relations
to effective temperatures for (decoupled) fast and slow dynamics.
In this work, the inclusion of correlations will be seen
to lead to an effective temperature~$\eps$,
related now to the {\it ratio} of these
two noises, in the slow dynamical regime.

When the material is weakly dilatant ($c\gg a$),
so that density fluctuations decay quickly to zero (and hence can be neglected),
the angle of repose $\theta(t)$ itself obeys an Ornstein-Uhlenbeck equation.
It relaxes exponentially fast to an equilibrium state, whose variance
\beq
\theta_\eq^2=\frac{\D_1^2}{a}
\eeq
is just the zero-dilatancy variance of $\theta$.

The opposite limit where $c\ll a$ is of much greater interest.
Here, density fluctuations are long-lived.
When, additionally, $\D_1$ is small, the angle
of repose has a {\it slow} dynamics reflective of the slowly evolving
density fluctuations.
These conditions are written more precisely as
\beq
\g\ll1,\qquad\eps\ll1,
\label{regime}
\eeq
in terms of two dimensionless parameters:
\beq
\g=\frac{c}{a},\qquad
\eps=\frac{ac^2\D_1^2}{b^2\D_2^4}=\frac{\theta_\eq^2}{\theta_\R^2}.
\label{regdef}
\eeq
The second equality follows from~(\ref{thesca}).
We see from this that $\eps$ is essentially the ratio
between the fluctuations of
the angle of repose $\theta$, in the respective limits
when it is decoupled from, and coupled to, the density
fluctuations $\phi$.
Giving $\eps$ the interpretation of an effective temperature (see below),
we define the regime~(\ref{regime})
as a {\it low-temperature and strongly dilatant} regime,
governed by the slow evolution of density fluctuations.

\section{Dynamics in and out of equilibrium}

Suppose that a sandpile is created in regime~(\ref{regime}) with very large
initial values for the angle $\theta_0$ and dilatancy $\phi_0$.
The initial stage of the dynamics is a transient one; here,
the noises are negligible,
so that the decay is entirely given by the deterministic parts
of~(\ref{dy1}) and~(\ref{dy2}):
\bea
&&\theta(t)=(\theta_0-\theta_\m)\e^{-at}+\theta_\m\,\e^{-2ct},\\
&&\phi(t)=\phi_0\,\e^{-ct},
\label{decay}
\eea
with
\beq
\theta_\m\approx\frac{b\,\phi_0^2}{a}.
\label{thmax}
\eeq
Thus,
density fluctuations $\phi(t)$ relax ex\-po\-nen\-tially,
while the trajectory $\theta(t)$ has two separate modes of relaxation:

\noindent $\bullet$
a fast (inertial) decay in $\theta(t)\approx\theta_0\,\e^{-at}$,
until $\theta(t)$ is of the order of $\theta_\m$,

\noindent $\bullet$
a slow (collective) decay in $\theta(t)\approx\theta_\m\,\e^{-2ct}$.

When $\phi(t)$ and $\theta(t)$
are small enough [i.e., $\phi(t)\sim\phi_\eq$
and $\theta(t)\sim\theta_\R$, cf.~(\ref{phisca}) and~(\ref{thesca})]
for the noises to have an appreciable effect, the above analysis
is no longer valid.
The system then reaches the equilibrium state
of the full non-linear stochastic process
represented by~(\ref{dy1}) and~(\ref{dy2}).
An exact analytical investigation of this,
for all values of the parameters~$\g$ and~$\eps$,
is presented in the Appendix.

In order to get a feeling for the more qualitative features
of the equilibrium state,
we note first that $\phi(t)$ is an Ornstein-Uhlenbeck process
with equilibrium variance:
\beq
\phi_\eq^2=\frac{\D_2^2}{c}.
\label{phisca}
\eeq
We see next that to a good approximation,
the angle $\theta$ adapts instantaneously to the dynamics of $\phi(t)$
in regime~(\ref{regime}):
\beq
\theta(t)\approx\frac{b\,\phi(t)^2}{a}.
\label{thinst}
\eeq
The two above statements together imply that
the distribution of the angle~$\theta(t)$ is approximately
that of the square of a Gaussian variable.
The typically {\it observed} angle of repose $\theta_\R$
is the time-averaged value
\beq
\theta_\R=\mean{\theta}_\eq=\frac{b\,\phi_\eq^2}{a}=\frac{b\D_2^2}{ac}.
\label{thesca}
\eeq
Equation~(\ref{thinst}) then reads
\beq
\theta(t)\approx\theta_\R\,\frac{\phi(t)^2}{\phi_\eq^2}.
\label{thred}
\eeq
Equation~(\ref{thred})
makes the physics behind the multivalued and history-dependent
nature of the angle of repose (referred to in the Introduction)
rather clear.
Its instantaneous value depends directly
on the instantaneous value of the dilatancy; its maximal (stable) value
$\theta_\m$ is noise-independent
[cf.~(\ref{thmax})] and depends only on the maximal
value of dilatancy that a given material can sustain stably\footnote{Here
stability is defined for times of order
$1/c$,
assumed to be
extremely~long.}.
Sandpiles constructed above this
will first decay quickly
to it; they will then decay more slowly to a `typical'
angle of repose~$\theta_\R$.
The ratio of these angles is given by
\beq
\frac{\theta_\m}{\theta_\R}=\frac{\phi_0^2}{\phi_\eq^2},
\label{thth}
\eeq
so that $\theta_\m\gg\theta_\R$ for $\phi_0\gg\phi_\eq$.
Since
spontaneous flow {\it always} occurs above $\theta_\m$,
it is known as the {\it angle of maximal stability}~\cite{br}.

Below this, i.e., for $\theta_\R<\theta<\theta_\m$,
we have a region of {\it bistability} which depends strongly
on sandpile history.
The above analysis shows that:

\noindent $\bullet$
Sandpiles submitted to {\it low} noise are {\it stable}
in this range of angles, at least for long times $\sim 1/c$.

\noindent $\bullet$
Sandpiles submitted to {\it high } noise [such that the effects
of dilatancy become negligible in~(\ref{dy1})]
{\it continue to decay rapidly} in this range of angles,
becoming nearly horizontal at short times $\sim 1/a$.

This provides rather satisfying agreement
with earlier work~\cite{bistability} on tilted sandpiles.
In that work, model sandpiles were submitted to tilts of
varying magnitudes before being restored
to their original (horizontal) state.
Large tilting resulted in avalanching,
while small tilting did not, even when the post-tilt
state involved the {\it same} angle of repose.
(The reason for this involved the very different
effects in each case on granular configurations
in the uppermost layers of the sandpile~\cite{bistability}.)
Bistability at the angle of repose was thus clearly manifest.

Our conclusions are that bistability at the angle
of repose is a natural consequence of applied noise
(tilt~\cite{bistability} or vibration) in granular systems;
a sandpile can either be at rest, or in motion,
at the {\it same} angle of repose, depending on its history.

\section{The dynamics of sandpile collapse}

\begin{figure}[htb]
\begin{center}
\includegraphics[angle=90,width=.6\linewidth]{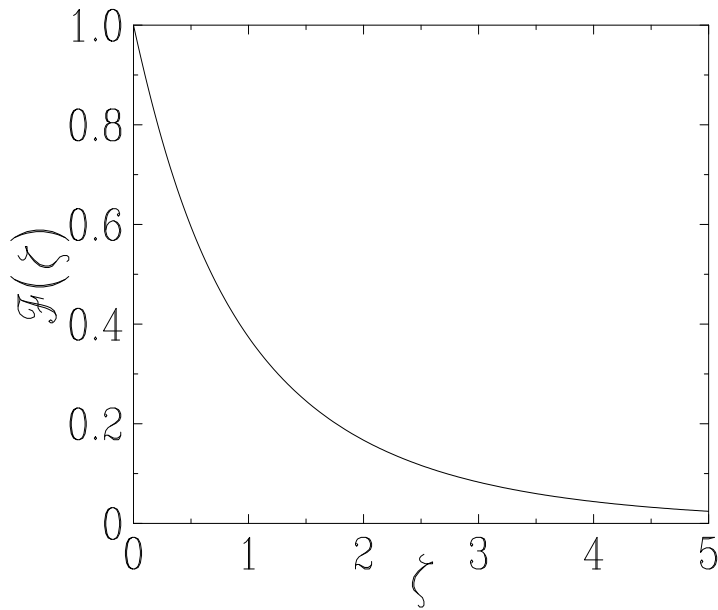}
\caption{\small
Plot of the function $\F(\zeta)$ entering the scaling law~(\ref{pitext})
of the probability $\Pi$ in the regime~(\ref{regime}).}
\label{figfzeta}
\end{center}
\end{figure}

We now examine the probability that under the prolonged
effect of low noise, the sandpile collapses; thus we look for events
where the angle $\theta(t)$ vanishes.
Such an event is expected to be very rare in the regime~(\ref{regime});
in fact it occurs only if the noise~$\eta_1(t)$
in~(\ref{dy1}) is sufficiently negative
for sufficiently long to compensate for the strictly positive term~$b\phi^2$.
The fully analytical confirmation of this argument
is presented in the Appendix.
It predicts that the
equilibrium probability for $\theta$ to be negative:
\beq
\Pi=\prob(\theta<0),
\eeq
scales throughout regime~(\ref{regime}) as
\beq
\Pi\approx\frac{(2\eps)^{1/4}}{\Gamma(1/4)}\,\F(\zeta),\qquad
\zeta=\frac{\g}{\eps^{1/2}}=\frac{b\D_2^2}{a^{3/2}\D_1}
\label{pitext}
\eeq
[see~(\ref{pisca})].
The scaling function $\F(\zeta)$, given explicitly in~(\ref{fzeta}),
decays mo\-no\-to\-ni\-cally from $\F(0)=1$ to $\F(\infty)=0$ (see
Figure~\ref{figfzeta}).

We explore further the regime $\zeta\gg1$,
where the `temperature' $\eps$ is much lower than the `barrier height' $\g^2$.
Here, the result~(\ref{fzas}) implies that the equilibrium probability
of collapse vanishes exponentially fast:
\beq
\Pi\sim\exp\left(-\frac{3}{2}\left(\frac{\g^2}{\eps}\right)^{1/3}\right).
\label{act}
\eeq
Thus complete collapse becomes an {\it activated} process:
collapse events occur at Poissonian times,
with an exponentially large characteristic time given by an Ar\-rhe\-nius~law:
\beq
\tau\sim1/\Pi
\sim\exp\left(\frac{3}{2}\left(\frac{\g^2}{\eps}\right)^{1/3}\right).
\label{arrhe}
\eeq

The stretched exponential in~(\ref{act}) is intriguing,
as it involves a fractional power
of the usual barrier-height-to-temperature ratio $\g^2/\eps$, and
therefore raises questions about the nature of the barriers involved.
We investigate this further below.

The equilibrium probability for the magnitude $\abs{\phi(t)}$
of the Ornstein-Uhlenbeck process
to remain smaller than some $\phi_1\ll\phi_\eq$
during a time interval $T$, is known to fall off exponentially~as
\beq
p_{\phi_1}(T)\sim\exp\left(-\frac{\pi^2}{4}
\,\frac{\phi_\eq^2}{\phi_1^2}\,cT\right).
\label{phismall}
\eeq
We now ask the question: what is the zero-temperature ($\eps=0$)
equilibrium probability $P_0(\theta)$ for the angle of repose to assume
a very small value $\theta\ll\theta_\R$?
Inserting $\phi_\eq^2/\phi_1^2\sim\theta_\R/\theta$ [cf.~(\ref{thred})]
in the estimate~(\ref{phismall}), with
$T\sim1/a$ (the relaxation time of the uncoupled $\theta$ dynamics),
we find that this probability is exponentially small:
\beq
P_0(\theta)\sim\exp\left(-k\frac{\g\theta_\R}{\theta}\right).
\label{pz}
\eeq
This somewhat heuristic argument is borne out by the rigorous
analysis of the Appendix, which leads to $k=1$,
and also predicts the prefactor~[see~(\ref{rhosca})].

In the presence of a low noise intensity $(\eps\ll1)$,
the exponential tail~(\ref{pz})
gets convoluted with a narrow Gaussian generated by the noise,
whose variance is $\theta_\eq^2=\eps\theta_\R^2$.
This gives
\beq
P(\theta)\sim\int_0^\infty\exp\left(-\frac{\g\theta_\R}{\theta_1}
-\frac{(\theta-\theta_1)^2}{2\eps\theta_\R^2}\right)\,\d\theta_1.
\eeq
Setting $\theta=0$, we get
\beq
\Pi\sim\int_0^\infty\exp\left(-\frac{\g\theta_\R}{\theta_1}
-\frac{\theta_1^2}{2\eps\theta_\R^2}\right)\,\d\theta_1.
\label{int1}
\eeq
The saddle-point of the above integral,
\beq
\theta_1\approx(\g\eps)^{1/3}\theta_\R,
\label{com1}
\eeq
leads to the stretched exponential probability distribution~(\ref{act}).

The above suggests a very strong analogy with
the famous problem of random trapping~\cite{trapping}.
Consider a particle performing Brownian motion in one dimension,
with diffusion constant $D$,
amidst a concentration $c$ of Poisson-distributed traps;
once a trap is reached, the particle ceases to exist.
The survival probability $S(t)$ of the particle
is therefore the probability that it has not encountered a trap until time $t$.
Assuming a uniform distribution of starting points,
the fall off of this probability can be estimated by
first computing the probability of finding a large region of length $L$ without
traps, and then weighing this with the probability that
a Brownian particle survives within it for a long time~$t$:
\beq
S(t)\sim\int_0^\infty\exp\left(-cL-\frac{\pi^2Dt}{L^2}\right)\,\d L.
\label{int2}
\eeq
The first exponential factor $\exp(-cL)$ is the probability that a
region of length $L$ is free of traps, whereas the second exponential factor
is the asymptotic survival probability of a Brownian particle
in such a region\footnote{This probability is $\exp(-Dq^2t)$, with
$q=\pi/L$ coming from the Dirichlet boundary conditions at
the absorbing endpoints.}.
The integral is dominated by a saddle-point at
\beq
L\approx\left(\frac{2\pi^2Dt}{c}\right)^{1/3},
\label{com2}
\eeq
whence we recover the well-known estimate
\beq
S(t)\sim\exp\left(-\frac{3}{2}\left(2\pi^2c^2Dt\right)^{1/3}\right).
\label{sur}
\eeq

We now elucidate as fully as possible the nature of the analogy,
in order to get further physical insight into
the problem of sandpile collapse.
Table~\ref{tableone} shows the quantitative correspondence between
parameters in both situations.
Both the survival probability $S(t)$ and the collapse probability $\Pi$
obey the stretched exponential laws~(\ref{sur}) and~(\ref{act}),
with an identical exponent $1/3$.
Both these anomalous dynamical laws
are the result of a saddle-point approximation,
which represents in each case an optimisation procedure.
In the trapping problem, large regions without traps are improbable,
whereas the particle would decay too fast in small ones
(it would get absorbed at a boundary);
the best compromise~(\ref{com2}) is found to scale as $L\sim t^{1/3}$.
In the sandpile problem, angles too far below
$\theta_\R$ are hard to find, as dilatancy
would resist their existence [cf.~(\ref{pz})]; on the other
hand, angles that are too large would inhibit collapse, given their larger
configurational barriers in the face of the noise.
Once again, the best compromise~(\ref{com1})
scales as $\theta_1\sim\eps^{1/3}$.

\begin{table}[ht]
\begin{center}
\begin{tabular}{|l|c|c|}
\hline
Problem&trapping&sandpile collapse\\
\hline
Physical quantity&$S(t)$&$\Pi$\\
Large parameter&$2\pi^2Dt$&$1/\eps$\\
Control parameter&$c$&$\g$\\
\hline
Integration variable&$L$&$\theta_\R/\theta_1$\\
Saddle-point value&$(2\pi^2Dt/c)^{1/3}$&$(\g\eps)^{-1/3}$\\
\hline
Result&(\ref{sur})&(\ref{act})\\
\hline
\end{tabular}
\caption{\small Quantitative correspondence between the derivations
of the stretched exponential laws~(\ref{sur}) and~(\ref{act})
in the one-dimensional random trapping and sandpile collapse problems.}
\label{tableone}
\end{center}
\end{table}

We use this analogy to develop the following picture for sandpile collapse.
Imagine that the collapse is visualised as the motion of an effective
particle (an exciton, say),
represented by the collective co-ordinate $\theta(t)$.
Under the influence of a temperature $\eps$
the exciton diffuses across a rough landscape
defined by the $\phi$ excitations; we can consider
this landscape to be a frozen background,
since the decay rate of $\theta$ is much faster than that of $\phi$.
Valleys are separated by $\phi$ barriers whose typical height scales as~$\g$;
however, sandpile collapse actually involves the traversal of a much lower
optimal barrier, given by~(\ref{act}).
The process of sandpile collapse
can therefore be visualised as the $\theta$ exciton's search for, and escape
across, the rare low barrier~(\ref{act}), in a frozen landscape
of large~$\phi$ barriers of typical height~$\g$.

At a given temperature $\eps$, sandpile collapse clearly
depends on the nature of the density fluctuations.
We look at two opposite cases of non-Gaussianness.
If density fluctuations are peaked around zero
(i.e., the material is almost non-dilatant),
this implies a much flatter, more ordered $\phi$-configurational
landscape, easier for the exciton to traverse.
This would lead to a `liquid-like' scenario of frequent collapse,
where a finite angle of repose would be hard to sustain under any
circumstances.
An explicit example is provided by the $\g\to0$ limit,
where the collapse probability scales as $\eps^{1/4}$~[see~(\ref{pitext})].

In the opposite case of strong dilatancy
(where large values of $\phi$ are more frequent than in the Gaussian
equilibrium distribution),
sandpile collapse is even more strongly inhibited.
If, for example, $\abs{\phi(t)}$
is constrained to remain larger than some threshold $\phi_\thr$,
the stretched exponential in~(\ref{act}) reverts
(in the $\eps\ll1$ regime considered) to an Arrhenius law in its usual form:
\beq
\Pi\sim\exp\left(-\frac{(\phi_\thr/\phi_\eq)^4}{2\eps}\right).
\eeq
This would arise in the case of a strongly dilatant material,
such as wet sand; angles of repose for such materials
can be far steeper than usual, and still resist collapse.

\section{Discussion}

In the above, we have looked at a very familiar problem,
that of the decay and eventual collapse of the angle of repose,
using an approach that combines new ideas with very traditional
concepts such as dilatancy~\cite{reynolds}.
Our simple theory suggests that sandpiles created
at arbitrarily large angles will decay quickly
to the maximal angle of stability; their subsequent behaviour
is bistable, with jamming at a typical finite angle of repose
as one outcome, or a continuing fast decay to zero, as another.
All of this occurs because of dynamical competition
between the fast dynamics of angle decay
and the slow dynamics of density fluctuations, especially
in the low-noise regime.
The collapse of a sandpile in the jammed regime
is shown to be a rare event; we have obtained exact results
for the ensuing activated process, which turns out to have
interesting analogies
with the well-known trapping problem.
Using these analogies, we are able to summarise
the process of collapse
as follows:
weakly dilatant sandpiles collapse easily, while
strongly dilatant ones bounce back.

\ack

AM warmly thanks SMC-INFM
(Research and Development Center for Statistical Mechanics and Complexity,
Rome, Italy), and the Service de Physique Th\'eorique, CEA~Saclay,
where parts of this work were done.
Kirone Mallick is gratefully acknowledged for interesting discussions,
and for having drawn our attention to Reference~\cite{luczka}
after the present investigation was completed.

\appendix

\section*{Appendix.
Equilibrium probability distribution for the process~(\ref{dy1}),~(\ref{dy2})}
\def\theequation{A.\arabic{equation}}

In this Appendix we derive an exact expression for the double Laplace transform
of the joint equilibrium probability distribution $P(\theta,\phi)$
of the variables $\theta$ and $\phi$,
for arbitrary values of the dimensionless parameters $\eps$ and $\g$,
and especially in the regime~(\ref{regime}).

Our starting point consists in writing down
the stationary (i.e., time-independent) Fokker-Planck equation
describing the equilibrium state
of the two-dimensional Markov process~(\ref{dy1}),~(\ref{dy2}).
This equation reads~\cite{vank}
\beq
\frac{\dpar J_\theta}{\dpar\theta}+\frac{\dpar J_\phi}{\dpar\phi}=0,
\eeq
where
\beq
J_\theta=-(a\theta-b\phi^2)P-\D_1^2\frac{\dpar P}{\dpar\theta},
\qquad J_\phi=-c\phi P-\D_2^2\frac{\dpar P}{\dpar\phi},
\eeq
i.e.,
\beq
\D_1^2\frac{\dpar^2P}{\dpar\theta^2}+\D_2^2\frac{\dpar^2P}{\dpar\phi^2}
+(a\theta-b\phi^2)\frac{\dpar P}{\dpar\theta}+c\phi\frac{\dpar P}{\dpar\phi}
+(a+c)P=0.
\label{fpe}
\eeq

We introduce the dimensionless variables
\beq
\w\theta=\frac{\theta}{\theta_\R},\qquad\w\phi=\frac{\phi}{\phi_\eq},
\eeq
where
\beq
\phi_\eq^2=\frac{\D_2^2}{c},\qquad\theta_\R=\frac{b\D_2^2}{ac}
\eeq
have been introduced in~(\ref{phisca}) and~(\ref{thesca}),
and define the double Laplace transform of $P(\theta,\phi)$ as
\beq
L(x,y)=\mean{\exp(-x\w\theta-y\w\phi)}
=\int\!\!\!\int\exp(-x\w\theta-y\w\phi)\,P(\theta,\phi)\,\d\theta\,\d\phi.
\eeq
In terms of this function, the Fokker-Planck equation~(\ref{fpe}) becomes
\beq
x\left(\frac{\dpar^2L}{\dpar y^2}+\frac{\dpar L}{\dpar x}\right)
=\eps x^2L+\g y\left(yL-\frac{\dpar L}{\dpar y}\right),
\label{lap}
\eeq
together with the normalisation $L(0,0)=1$.
The dimensionless parameters
\beq
\g=\frac{c}{a},\qquad\eps=\frac{ac^2\D_1^2}{b^2\D_2^4},
\eeq
have been introduced in~(\ref{regdef}).

Our main interest will reside in the distribution of $\w\theta$,
encoded in the Laplace transform
\beq
f(x)=L(x,0)=\mean{\exp(-x\w\theta)},
\eeq
so that
\beq
P(\w\theta)=\int\frac{\d x}{2\imi\pi}\,\e^{x\w\theta}\,f(x),
\label{rhoint}
\eeq
and especially in the probability for $\theta$ to be negative:
\beq
\Pi=\prob(\theta<0)
=\int_{-\infty}^0P(\w\theta)\,\d\w\theta
=\int\frac{\d x}{2\imi\pi}\,\frac{f(x)}{x}.
\label{piint}
\eeq

We first notice that
\beq
L(0,y)=\e^{y^2/2}
\eeq
obeys~(\ref{lap}), in agreement with the plain observation
that the stationary distribution of $\w\phi$ is a Gaussian with unit variance.
The full problem can be solved by making the Ansatz
that $L(x,y)$ keeps the same functional form in $y$ for any fixed value of $x$,
i.e., looking for a solution to~(\ref{lap}) of the form
\beq
L(x,y)=f(x)\,\exp\left(g(x)\,\frac{y^2}{2}\right),
\label{ansatz}
\eeq
with $f(0)=g(0)=1$, where $g(x)$ is the `$x$-dependent variance' of $\phi$.
Equation~(\ref{lap}) boils down to two
ordinary differential equations for $f(x)$ and $g(x)$:
\bea
&&x(g'+2g^2)+2\g(g-1)=0,\label{eqg}\\
&&f'=(\eps x-g)f\label{eqf},
\eea
justifying thus the validity of the above Ansatz.

The above equations can be solved as follows.
Equation~(\ref{eqg}) is a Riccati equation~\cite{ode} for $g(x)$,
which can be linearised by setting
\beq
g=\frac{1}{2}\left(\frac{\psi'}{\psi}-\frac{\g}{x}\right).
\label{gric}
\eeq
The new unknown function $\psi(x)$ obeys the second-order linear equation
\beq
\psi''=\left(\frac{\g(\g-1)}{x^2}+\frac{4}{x}\right)\psi,
\eeq
whose normalised regular solution reads
\beq
\psi(x)=x^\g\sum_{n\ge0}\frac{\Gamma(2\g)}{\Gamma(n+2\g)}
\,\frac{(4\g x)^n}{n!}
=\frac{\Gamma(2\g)}{(4\g)^\g}\,
2\sqrt{\g x}\,I_{2\g-1}(4\sqrt{\g x}),
\label{psi}
\eeq
where $I_{2\g-1}$ is the modified Bessel function.
Equations~(\ref{gric}) and~(\ref{eqf}) then respectively yield
the explicit expressions
\beq
g(x)=\sqrt{\frac{\g}{x}}
\,\frac{I_{2\g}(4\sqrt{\g x})}{I_{2\g-1}(4\sqrt{\g x})}
\label{gres}
\eeq
and
\beq
f(x)=f_0(x)\,\e^{\eps x^2/2},
\label{fres}
\eeq
with
\bea
f_0(x)=\left(\frac{x^\g}{\psi(x)}\right)^{1/2}
&=&\left(\sum_{n\ge0}\frac{\Gamma(2\g)}{\Gamma(n+2\g)}\,\frac{(4\g x)^n}{n!}
\right)^{-1/2}\nonumber\\
&=&\left(\frac{(4\g x)^{\g-1/2}}
{\Gamma(2\g)\,I_{2\g-1}(4\sqrt{\g x})}\right)^{1/2}.
\label{f0res}
\eea

The product formula~(\ref{fres}) for $f(x)$
expresses that the variable $\w\theta$
is the convolution of two independent random variables:

\noindent (i)
a positive random variable $\w\theta_0$ whose distribution,
encoded in $f_0(x)$, only depends on~$\g$;

\noindent (ii)
a Gaussian variable with variance $\eps$.

The cumulants $c_n$ of $\w\theta$ can be obtained by means of
the series expansion
\beq
\sum_{n\ge1}\frac{c_n(-x)^n}{n!}=\ln f(x)
=\frac{\eps x^2}{2}-\frac{1}{2}
\ln\sum_{n\ge0}\frac{\Gamma(2\g)}{\Gamma(n+2\g)}\,\frac{(4\g x)^n}{n!}.
\eeq
We thus get
\bea
c_1=1,\quad c_2=\frac{2}{2\g+1}+\eps,\quad c_3=\frac{8}{(\g+1)(2\g+1)},
\nonumber\\
c_4=\frac{48(5\g+3)}{(\g+1)(2\g+1)^2(2\g+3)},\quad
c_5=\frac{384(7\g+6)}{(\g+1)(\g+2)(2\g+1)^2(2\g+3)},
\nonumber\\
c_6=\frac{3840(42\g^3+118\g^2+107\g+30)}
{(\g+1)^2(\g+2)(2\g+1)^3(2\g+3)(2\g+5)},
\eea
and so on.
All the cumulants of $\w\theta$ coincide with those of $\w\theta_0$,
except $c_2$, whose term linear in $\eps$ represents
the Gaussian variable in (ii) above.

Let us first consider the limiting situation $\g=0$,
which is reached when the rate $c$ vanishes,
so that the dynamics of the variable $\phi$ is entirely frozen.
The above solution simplifies greatly.
Indeed only the terms corresponding to $n=0$ and $n=1$
survive in the series representation~(\ref{psi}) for~$\psi(x)$, so that
\beq
\psi(x)=1+2x,
\eeq
and therefore
\beq
f_0(x)=(1+2x)^{-1/2},\qquad g(x)=\frac{1}{1+2x}.
\label{prod}
\eeq
The expression for $f_0(x)$ shows that $\w\theta_0$
is nothing but the square of a normalised Gaussian variable,
the latter being identified with $\w\phi$.

The probability $\Pi$ for $\theta$ to be negative therefore reads
\beq
\Pi=\int\!\frac{\d x}{2\imi\pi}\,\frac{\e^{\eps x^2/2}}{x\sqrt{1+2x}}.
\eeq
In the regime $\eps\ll1$ of most interest, this expression simplifies as
\beq
\Pi\approx\int\!\frac{\d x}{2\imi\pi}\,\frac{\e^{\eps x^2/2}}{\sqrt{2x^3}}
\approx\frac{(2\eps)^{1/4}}{\Gamma(1/4)}.
\label{pizero}
\eeq

In the general situation where $\eps$ and $\g$ are both non-zero,
the results~(\ref{gres})--(\ref{f0res}) are much more involved.
The following situations deserve our attention:

\smallskip\noindent $\bullet$
For $\eps=0$, i.e., in the absence of noise in~(\ref{dy1}),
the variable $\w\theta$ reduces to $\w\theta_0$.
Its probability density reads
\beq
P_0(\w\theta)=\int\!\frac{\d x}{2\imi\pi}\,\e^{x\w\theta}
\,\left(\frac{(4\g x)^{\g-1/2}}
{\Gamma(2\g)\,I_{2\g-1}(4\sqrt{\g x})}\right)^{1/2}.
\eeq
The parenthesis falls off exponentially for $x\to+\infty$,
confirming thus that $\w\theta$ is positive.
As $\w\theta\to0$, the integral in the right-hand side
is dominated by large values of $x$,
where the Bessel function can be approximated by its asymptotic expression
\beq
I_{2\g-1}(z)\approx\frac{e^z}{\sqrt{2\pi z}},
\eeq
irrespective of $\g$.
A saddle-point approximation yields the estimate
\beq
P_0(\w\theta)\approx
\frac{\left(2\g/\w\theta\right)^\g}{\Gamma(2\g+1)^{1/2}}
\;\left(\frac{8\g^3}{\pi\w\theta^5}\right)^{1/4}
\exp\left(-\frac{\g}{\w\theta}\right).
\label{rhosca}
\eeq
The probability density of $\w\theta$ therefore falls off exponentially fast
for $\w\theta\ll\g$.

\smallskip\noindent $\bullet$
For $\eps>0$, the probability for $\w\theta$ to be negative,
\beq
\Pi=\int\!\frac{\d x}{2\imi\pi x}\,\e^{\eps x^2/2}
\left(\frac{(4\g x)^{\g-1/2}}
{\Gamma(2\g)I_{2\g-1}(4\sqrt{\g x})}\right)^{1/2},
\eeq
is non-zero.
As $\eps\to0$, the integral in the right-hand side
is again dominated by large values of $x$,
and a saddle-point approximation yields the estimate
\beq
\Pi\approx
\frac{\left(8\g^2/\eps\right)^{\g/3}}{\Gamma(2\g+1)^{1/2}}
\;\left(\frac{8\eps}{9\pi}\right)^{1/4}
\exp\left(-\frac{3}{2}\left(\frac{\g^2}{\eps}\right)^{1/3}\right).
\label{piexp}
\eeq
The probability $\Pi$ therefore falls off as a stretched exponential
for $\eps\ll\g^2$.

\smallskip\noindent $\bullet$
In the $\g\to0$ limit,
the results~(\ref{rhosca}) and~(\ref{piexp}) simplify,
as the first factors of both expressions go to unity.
Furthermore, throughout the regime~(\ref{regime}),
i.e., whenever $\eps$ and~$\g$ are simultaneously small,
the probability $\Pi$ obeys a scaling law of the form
\beq
\Pi\approx\frac{(2\eps)^{1/4}}{\Gamma(1/4)}\,\F(\zeta),\qquad
\zeta=\frac{\g}{\eps^{1/2}}.
\label{pisca}
\eeq
This result interpolates between~(\ref{pizero}) and~(\ref{piexp}).
In order to derive the expression of the scaling function $\F(\zeta)$,
it is again convenient to come back to the series
representation~(\ref{psi}) for~$\psi(x)$.
For $\g\ll1$, keeping $z=4\sqrt{\g x}$ finite,
that expression simplifies as
\beq
\psi(x)\approx\frac{z\,I_1(z)}{4\g}.
\eeq
Inserting this scaling estimate into~(\ref{fres}) and~(\ref{piint}),
we obtain after some algebra
\beq
\F(\zeta)=2^{7/4}\,\Gamma(1/4)\,\zeta^{1/2}
\int_{\e^{-\imi\pi/4}\infty}^{\e^{\imi\pi/4}\infty}
\frac{\d z}{2\imi\pi}\left(z^3I_1(z)\right)^{-1/2}
\exp\left(\frac{z^4}{512\,\zeta^2}\right).
\label{fzeta}
\eeq
This is a monotonically decreasing function of $\zeta$,
starting from the value $\F(0)=1$, so that~(\ref{pizero}) is recovered.
Its fall-off at large values of $\zeta$,
\beq
\F(\zeta)\approx\Gamma(1/4)\,\left(\frac{4}{9\pi}\right)^{1/4}
\exp\left(-\frac{3}{2}\,\zeta^{2/3}\right),
\label{fzas}
\eeq
agrees with~(\ref{piexp}) in the $\g\to0$ limit.

\end{document}